\documentclass{article}
\usepackage{graphicx}
\usepackage{amsmath}
\usepackage{amssymb}

\title{Polarized Spacetime Foam}
\author{V. Dzhunushaliev
\thanks{E-mail : dzhun@hotmail.kg}}

\date{}

\begin{document}
\maketitle

\begin{center}
\textit{
Dept. Phys. and Microelectronics Engineer., Kyrgyz-Russian
Slavic University, Bishkek, Kievskaya Str. 44, 720000, Kyrgyz
Republic}
\end{center}

\begin{abstract}
An approximate model of a spacetime foam is presented. It is supposed that 
in the spacetime foam 
each quantum handle is like to an electric dipole and therefore the spacetime 
foam is similar to a dielectric. If we neglect of linear sizes of the quantum 
handle then it can be described with an operator containing a Grassman 
number and either a scalar or a spinor field. For both fields the Lagrangian 
is presented. For the scalar field it is the dilaton gravity + electrodynamics 
and the dilaton field is a dielectric permeability. The spherically 
symmetric solution in this case give us the screening of a bare electric 
charge surrounded by a polarized spacetime foam and the energy of the electric 
field becomes finite one. In the case of the spinor field the spherically symmetric 
solution give us a ball of the polarized spacetime foam filled with the confined 
electric field. It is shown that the full energy of the electric field in the ball 
can be very big. 
\end{abstract}

\section{Introduction} 

One of the manifestation of quantum gravity is a spacetime foam introduced 
by Wheeler \cite{wheel1} for the description a hypothesized topology fluctuations 
on the Planck 
scale level. The spacetime foam is a cloud of appearing/disappearing quantum 
handles. The appearance/desctruction of these handles leads to the change of 
spacetime topology. This fact give rise to big difficulties at the description 
of the spacetime foam since by topology changes of a space (according to 
Morse theory \cite{dubrovin} (Part2, Chap. 2, Sec. 10)) 
the critical points must exist where the time direction is not defined. 
In each such point should be a singularity which is an obstacle for the mathematical 
description of the spacetime foam.
\par 
Nevertheless we can try to describe spacetime foam by some approximate manner. 
For the beginning we offer a model of the single quantum handle as the wormhole-like 
solution in the Kaluza-Klein gravity. In some approximation we can neglect the 
linear sizes of the handle and in this case each quantum handle looks as two 
points pasted together. It 
is so called a minimalist wormhole \cite{smolin}, \textit{i.e.} quantum wormhole 
in which the cross section of the throat is contracted to a point. Afterwards 
we will introduce an operator which describes the creation/annihilation of the 
minimalist wormhole. We will show that this operator can be presented as 
a Grassman number and some (scalar or fermion) field. 
\par
It is interesting to compare the model of spacetime foam presented here 
with another approaches in this field. Now we would like briefly 
to circumscribe these approaches in the investigations of spacetime foam 
(for details see review \cite{garay1}). 
The spacetime foam and the complicated topological structure 
connected with it may provide a mechanism for explaining the vanishing 
of the cosmological constant and for fixing all the 
constants of nature \cite{coleman88b}. The spacetime foam may 
also induce loss of a quantum coherence \cite{hawk}, 
and produce a frequency-dependent energy shifts that would slightly 
alter the dispersion relations for the different low-energy fields 
\cite{garay2,garay3}. Also spacetime foam has been proposed as a mechanism for 
regulating both the ultraviolet \cite{crane86} and the infrared 
behavior of quantum field theory \cite{magnon88}. The most of these 
investigations based on the properties of a single or $N$ wormholes 
forming spacetime foam. Here in the presented model we shall try to describe 
some effects connected to foam not being interested their internal structure. 
The difference between this model and above-mentioned approaches is similar 
to difference between thermodynamics and classical mechanics. 
The classical mechanics investigates the movement of each single particle 
in a gas whereas the thermodynamics works with physical quantities which 
are the average values of microscopical quantities. In the 
offered model of spacetime foam we are interesting only for cooperative 
properties of the foam that allows us to introduce some effective field 
which describes approximately and effectively spacetime foam. 
\par
One mouth of quantum wormhole can entrap the electric force lines and then these 
force lines will emerge from another mouth. It allows us to consider each 
quantum wormhole as an electric dipole and the spacetime foam as a dielectric 
\cite{dzh2}. In this case 
a big electric field can polarize the spacetime foam that can be the reason 
for some physical effects. 

\section{The model of a quantum handle}
\label{model}

\begin{figure}
\begin{center}
\framebox{
\includegraphics[height=5cm,width=10cm]{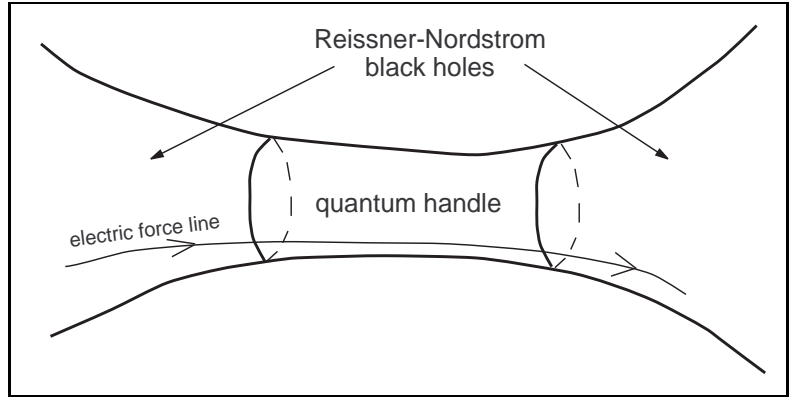}}
\vspace{5mm}
\caption{The model of a quantum handle.
The whole spacetime is 5 dimensional and: in the
Reissner-Nordstr\"om black hole $G_{55} = const$ and it is not
varying (this is the 5D gravity in the initial Kaluza-Klein
interpretation); in the 5D throat $G_{55}$ is the dynamical
field.}
\label{fig1}
\end{center}
\end{figure}

The model of the individual quantum handle is presented on the 
Fig.(\ref{fig1}). It is some realization of the Wheeler idea about
a wormhole entrapping electric force lines. In Ref.\cite{wheel1}
he wrote: ``Along with the fluctuations in the metric there occur
fluctuations in the electromagnetic field. In consequence
the typical multiply connected space $\ldots$ has a net flux of
electric lines of force passing through the "wormhole". These
lines are trapped by the topology of the space. These lines give
the appearance of a positive charge at one end of the wormhole
and a negative charge at the other''.
The composite wormhole on the Fig.(\ref{fig1}) consists
from two Reissner-Nordstr\"om black holes and the 5D throat
inserted between them \cite{dzh7}. The 5D metric for this throat
is
\begin{equation}
ds^{2}_{(5)} = - R_0^2 e^{2\psi(r)}\Delta (r)
\left(d\chi  + \omega (r)dt + Q \cos \theta d\varphi \right)^2 + 
\frac{1}{\Delta (r)}dt^{2} - dr^{2} - a(r)
\left (
d\theta ^2 + \sin\theta ^2 d\varphi ^2
\right ),
\label{ind1}
\end{equation}
where $\chi $ is the 5$^{th}$ extra coordinate; $R_0$ and $Q$ are some constants. 
We assume that in some rough approximation the quantum handle in the spacetime foam 
can be presented by this manner. The 5D Einstein equations for the throat are 
\begin{eqnarray}
  \frac{\Delta ''}{\Delta} - \frac{{\Delta'}^2}{\Delta^2} + 
  \frac{\Delta' \psi'}{\Delta} + \frac{a' \Delta'}{a \Delta} + 
  R_0^2{\omega'}^2 \Delta^2 e^{2\psi} & = & 0 ,
\label{ind1a}\\
  \omega '' + \omega'\left( 2\frac{\Delta'}{\Delta} + 3\psi' + 
  \frac{a'}{a}\right) & = & 0 ,
\label{ind1b}\\
  \frac{a''}{a}   + \frac{a'\psi'}{a} - \frac{2}{a} + 
  \frac{Q^2\Delta e^{2\psi}}{a^2} & = & 0 ,
\label{ind1c}  \\
  \psi'' + {\psi'}^2 + \frac{a'\psi'}{a} - 
  \frac{Q^2\Delta e^{2\psi}}{2a^2} & = & 0 ,
\label{ind1d}  \\
  \frac{{\Delta'}^2}{\Delta^2} + 2\frac{\Delta'\psi'}{\Delta} - 
  4\frac{a'\psi'}{a} + \frac{4}{a} - \frac{{a'}^2}{a^2} - 
  R_0^2{\omega'}^2 \Delta^2 e^{2\psi} - 
  \frac{Q^2\Delta e^{2\psi}}{a^2}& = & 0 .
\label{ind1e}  
\end{eqnarray}
It can be shown \cite{dzhsin} that there is three type of solutions : 
the first type (wormhole-like solution) is presented on Fig.(\ref{fig1}) 
with $E > H$ ($E$ and $H$ are Kaluza-Klein electric and magnetic fields), 
the second one is an infinite flux tube with $E = H$ and the third one 
is a singular solution (finite flux tube) with $E < H$. 
The longitudinal size $L_0$ of the WH-like solution depends on the 
relation between electric and magnetic fields : if $(1 - H/E) \rightarrow 0$ then 
$L_0 \rightarrow \infty$. Now we will describe some properties of the 
wormhole-like solutions allowing us to interpret their as quantum handles 
in spacetime foam.
\par 
Let us define an approximate solution 
close to points $r^2 = r^2_0$ (where $ds^2(\pm r_0) = 0$). 
\begin{eqnarray}
  \Delta (r)& \approx & \Delta_1 \left( r_0^2 - r^2 \right) , 
\label{ind1f}\\
  \omega (r)& \approx & \frac{\omega_1}{r_0^2 - r^2} ,
\label{ind1g}\\
  \psi (r)& \approx & \frac{\psi_3}{6} \left( r_0^2 - r^2 \right)^3 .
\label{ind1h}  
\end{eqnarray}
with  
\begin{eqnarray}
  \Delta_1 & = & \pm \frac{q}{2a_0r_0} , 
\label{ind1j}\\
  \omega_1 & = & \frac{2 a_0 r_0}{q} ,
\label{ind1k}\\
  \psi_3 & = & \pm \frac{qQ^2}{2a_0^3 r_0^3} 
\label{ind1l}  
\end{eqnarray}
here $a_0 = a(r=\pm r_0)$, $q$ is some constant. It is easy to show 
that at the hypersurfaces $r = \pm r_0$ : $ds^2 = 0$. On these 
hypersurfaces the change of the metric signature takes place : 
$(+,-,-,-,-)$ by $|r| < r_0$ and $(-,-,-,-,+)$ by 
$|r| > r_0$. Following to Bronnikov \cite{bron} we call these 
two hypersurfaces as $T-$horizons. 
\par 
For the definition of a Kaluza-Klein electric field we 
consider Eq.(\ref{ind1b}) 
\begin{equation}
  \left[\left(\omega' \Delta^2 e^{3\psi}\right) 
  4\pi a  
  \right]' = 0 
\label{ind1q}
\end{equation}
here $4\pi a$ is the area of $S^2$ sphere. Comparing with the Gauss law we see that 
Kaluza-Klein electric field $E_{KK}$ can be defined as follows 
\begin{equation}
  E_{KK} = \omega' \Delta^2 e^{3\psi} = \frac{q}{a}
\label{ind1w}
\end{equation}
here $q$ is an electric charge which is proportional to a flux of 
electric field. In this case the force lines of the electric field 
are uninterrupted and can be continued through the surfaces of matching 
the 5D WH-like solution and the Reissner-Nordstr\"om solution like to 
Fig.\ref{fig1}. 
\par
\begin{figure}
\begin{center}
\framebox{
\includegraphics[height=5cm,width=10cm]{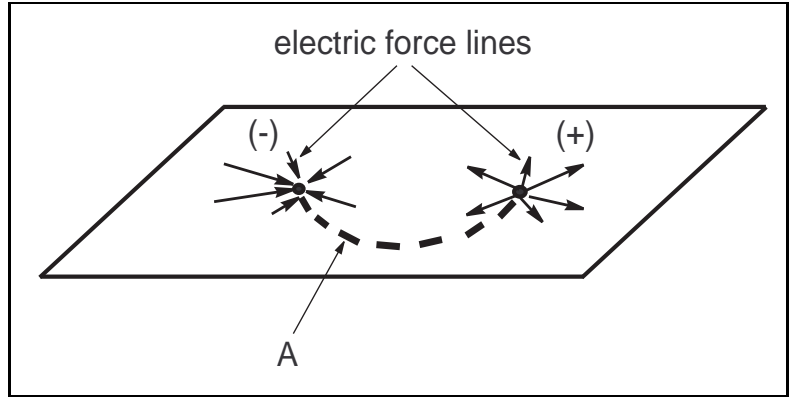}}
\vspace{5mm}
\caption{The minimalist wormhole as a quantum handle in the spacetime foam. 
One mouth entraps and another one emits 
the electric force lines and looks as two connected points. 
The curve \textbf{A} schematically shows the identification of these two 
points}
\label{fig1a}
\end{center}
\end{figure}

On these $T-$horizons we should match:
\begin{itemize}
\item
the flux of the 4D electric field (defined by the 4D Maxwell
equations) with the flux of the 5D electric field defined
by $R_{5t} = 0$ Kaluza-Klein equation.
\item
the area of the Reissner-Nordstr\"om event horizon with
the area of the $T-$horizon.
\end{itemize}
In the spacetime foam can be two different types of quantum handles : 
the first type does not entrap the electric force lines, the second type 
respectively entrap these lines. We will not consider the first type of quantum 
handles. In this case the quantum handle of the second type (in the minimalist 
wormhole approximation) is presented on Fig. \ref{fig1a}. 
\par 
In the next section we will consider an approximate model of the spacetime 
foam where we neglect the cross section of quantum handles. It means that 
the area of cross section of the throat is contracted into 
a point. Such approximation for the quantum handle is called as the minimalist 
wormhole. And further we will consider the consequences for such approximation. 
Thus, in this section we have presented a microscopic model 
of the quantum handle. But in the next sections we will 
consider an effective model of the spacetime foam where we will neglect 
of the cross section of quantum handles in the spacetime foam. 

\section{The operator description of the minimalist wormhole}

Let us introduce an operator $W(x,y)$ which describes creation/annihilation 
minimalist wormhole connected two points $x$ and $y$ \cite{smolin} \cite{dzh1}. 
Let the operator $W(x,y)$ 
has the following property 
\begin{equation}
  W^2(x,y) = 0.
\label{sec1-10}
\end{equation}
It means that the reiterated creation/annihilation minimalist wormhole is 
senseless. This property tell us that the operator $W(x,y)$ can be connected 
with the Grassman numbers. In the general case the operator $W(x,y)$ is nonlocal 
one but in this paper we will consider the simplest case when $W(x,y)$ can be factorized 
on two local operators 
\begin{equation}
  W(x,y) = \theta(x) \theta(y).
\label{sec1-20}
\end{equation}
The operator $\theta(x)$ can be considered as \textit{a readiness} of a point $x$ to 
pasting together with a point $y$ (or conversely the separation these points 
in the minimalist wormhole). 
\par 
We will consider three cases. 
\par
\textbf{\underline{The first case}} is
\begin{equation}
  W^{ab}(x,y) = \phi(x) \theta^a \phi(y) \theta^b
\label{sec1-30}
\end{equation}
where $a,b$ are the undotted spinor indices, $\phi(x)$ is the scalar field, 
$\theta^a$ is the Grassman number 
\begin{equation}
  \theta^a \theta^b + \theta^b \theta^a = 0.
\label{sec1-40}
\end{equation}
It is easy to proof that 
\begin{equation}
  W^{ab}(x,y) W_{ab}(x,y) = \phi^2(x) \phi^2(y) 
  \theta^a \theta^b \theta_a \theta_b = 
  \phi^2(x) \phi^2(y) \varepsilon_{aa'}\varepsilon_{bb'}
  \theta^a \theta^b \theta^{a'}\theta^{b'} \equiv 0
\label{sec1-50}
\end{equation}
where 
\begin{equation}
    \varepsilon^{ab} = 
    \begin{pmatrix}
    0 & 1 \\
    -1 & 0
  \end{pmatrix} ,
  \qquad 
  \varepsilon_{ab} = 
    \begin{pmatrix}
    0 & -1 \\
    1 & 0
  \end{pmatrix} .
\label{sec1-70}
\end{equation}

Eq. \ref{sec1-50} means that it makes no sense to identify two points 
twice. 
\par
\textbf{\underline{The second case}} is
\begin{equation}
  W(x,y) = \psi_a(x) \theta^a \psi_b(y) \theta^b = 
  \psi_a(x) \psi_b(y) \theta^a \theta^b = 
  \left(
  \psi_1(x) \psi_2(y) - \psi_2(x) \psi_1(y)
  \right)
  \theta^1 \theta^2
\label{sec1-80}
\end{equation}
where $\psi_a(x)$ is an undotted spinor field in 
$(\frac{1}{2},0)$ representation. The square is zero
\begin{equation}
  W^2(x,y) = \left(
  \psi_1(x) \psi_2(y) - \psi_2(x) \psi_1(y)
  \right)^2
  \theta^1 \theta^2 \theta^1 \theta^2 \equiv 0.
\label{sec1-90}
\end{equation}
In this case $W(x,x) \equiv 0$. It means that the minimalist wormhole can 
connect only two different points. 
\par
\textbf{\underline{The third case}} is
\begin{equation}
  W(x,y) = \psi_a(x) \theta^a \psi_{\dot b}(y) \bar{\theta}^{\dot b} = 
  \psi_a(x) \psi_{\dot b}(y) \theta^a \bar{\theta}^{\dot b}
\label{sec1-100}
\end{equation}
where $\psi_{\dot b}$ is a dotted spinor in $(0,\frac{1}{2})$ representation. 
The square is zero 
\begin{equation}
  W^2(x,y) = \left( 
  \psi_a(x) \psi_{\dot b}(y) \theta^a \bar{\theta}^{\dot b}
  \right)
  \left( 
  \psi_c(x) \psi_{\dot d}(y) \theta^c \bar{\theta}^{\dot d}
  \right) = 
  \Bigl(
  \psi_a(x) \psi_c(x) \theta^a \theta^c 
  \Bigr)
  \left(
  \psi_{\dot b}(y) \psi_{\dot d}(y) \bar{\theta}^{\dot b} \bar{\theta}^{\dot d}
  \right) \equiv 0
\label{sec1-110}
\end{equation}
as 
\begin{equation}
  \psi_a(x) \psi_c(x) \theta^a \theta^c = 
  \psi_1(x) \psi_2(x) \theta^1 \theta^2 + 
  \psi_2(x) \psi_1(x) \theta^2 \theta^1 \equiv 0.
\label{sec1-120}
\end{equation}
We must note that as the classical canonical theory cannot describes topology 
change these operators do not correspond to any classical observables. 
It corresponds to the well-known fact that the Grassman numbers do not have 
any classical interpretation. 
\par 
We can suppose that $W(x,y)$ operator is described with dynamical fields : 
scalar field $\phi(x)$ or spinor field $\psi(x)$. Below we will present some 
equations for these fields and discuss the physical sense of corresponding solutions. 
\par 
It is interesting to note that in Ref. \cite{garay2} very similar idea 
is considered 
that the gravitational interactions (topology fluctuations) 
presented in the spacetime foam are modelled by means of nonlocal interactions 
(operator $W(x,y)$ in our case) that relate spacetime points and these 
nonlocal interactions can be described in terms of local interactions 
($\theta(x)$ in our case).
\par 
The operator $W(x,y)$ can be connected with an indefiniteness 
(the loss of information) of our knowledge about two points $x^\mu$ and 
$y^\mu$~: we do not know if these points are connected by the 
quantum minimalist wormhole or not. We would like to note that the similar 
idea was considered in Ref. \cite{krasnov} : the author investigates 
the consequences of the loss of information about the topology of 
the background spacetime and found that the existence of spacetime foam 
leads to a situation where the number of bosonic fields is a variable 
quantity.  
\par 
Are the functions $\phi(x)$ and $\psi(a(x)$ dynamical or not ? Of coarse, 
in the context of this paper we do not have the answer on this question 
because here is proposed only an effective model of the spacetime foam. 
The exact model of the spacetime foam can be obtained only from a quantum 
gravity theory. From the physical point of view we can expect that the state 
and properties of spacetime foam depend on the magnitude of gravitational 
and external (for example, E\&M) fields. The aim of this paper is to try 
to describe the state of spacetime foam by some effective way using some 
(scalar/spinor) field. Common sense guides us to suppose that it is possible 
if we neglect the cross section of each quantum handle in the spacetime foam. 
It means that each quantum handle looks like to two points connected by a line 
(see, Fig. \ref{fig1a}) and this (scalar or spinor) field give us 
an information about quantum 
handles. Below we will see that it is the information about the typical length 
of quantum handles in the spacetime foam. 
\par 
Another question appearing in the context of this section is : what kind 
of wormhole operator is realized in the real spacetime foam or every 
such possibility can happen in the spacetime foam ? Again the answer lies in 
quantum gravity theory. Nevertheless, we can compare the results of similar 
calculations for the different choice of the minimalist operator in order 
that to select those coincides with our common sense. 
For this we will compare the wormhole operator $W(x,y)$ for the different cases : 
for the scalar and spinor fields. It is evident that 
$W(x,y) \approx 0$ if the points $x$ and $y$ are well away from each other. 
For example, for the ordinary spacetime foam (without any external influence) 
\begin{equation}
  W(x,y) \approx 0 
  \quad \text{for} \quad 
  |x - y| >l_{Pl} .
\label{sec1-130}
\end{equation}
Consequently we will use this criteria in the next sections for the choice 
between all possibilities for the effective field (scalar or spinor).

\section{The scalar field}

For the description of the dynamics of the scalar field we offer the following 
(dilaton gravity + Maxwell) Lagrangian 
\begin{equation}
  {\cal L} = -\sqrt{-g} 
  \left(
  R + 2\nabla_\mu \phi \nabla^\mu \phi - e^{-2\phi}
  F_{\mu\nu} F^{\mu\nu}
  \right)
\label{sec2-10}
\end{equation}
where $g$ is the determinant of the 4D metric, 
$R$ is the Ricci scalar, $\phi$ is the dilaton field and 
$F_{\mu\nu} = \partial_\mu A_\nu - \partial_\nu A_\mu$ 
is the tensor of the electromagnetic field, $\mu = 0,1,2,3$. 
It means that we interpret the 
above-mentioned scalar field as the dilaton field. The corresponding field 
equations are 
\begin{eqnarray}
  \nabla_\nu \left( e^{-2\phi}F^{\mu\nu} \right) & = & 0,
\label{sec2-20}\\
  \nabla_\mu \nabla^\mu \phi & = & \frac{1}{2} e^{-2 \phi} 
  F_{\mu\nu} F^{\mu\nu},
\label{sec2-30}\\
  R_{\mu\nu} - \frac{1}{2} R & = & \varkappa T_{\mu\nu}
\label{sec2-40}  
\end{eqnarray}
where $R_{\mu\nu}$ is the Ricci tensor and $T_{\mu\nu}$ is the 
tensor-momentum for dilaton and Maxwell fields. Each minimalist wormhole 
(as a quantum handle in the spacetime foam) looks as a dipole because one mouth 
entraps the electric force lines and another mouth will emerges these force lines. 
At first sight our choice of scalar field as the dilaton field is arbitrary but 
now we would like to explain why it is the natural choice. If we take a look 
over Fig.\ref{fig1a} than we see that the electric force lines around the 
quantum handle behave like to electric force lines in the presence of an 
electric dipole. How we can effectively (and approximately) describe the set 
of minimalist wormholes which entrap and emerge electric force lines ? We 
suppose that an external electric field acts on the minimalist wormhole 
(quantum handle) so that they line up with the electric field. Then such 
orientation of the minimalist wormholes will change the external electric field 
just as it happens in a dielectric. Another words we can introduce an effective 
electric displacement $\vec D$ which will describe the result of polarization 
of spacetime foam. As usually this vector $\vec D$ should be connected with 
the electric field as $\vec D = \varepsilon \vec E$ (where $\varepsilon$ is a 
dielectric permeability). The ordinary Maxwell equations looks as 
\begin{equation}
  \partial_\nu \left( \varepsilon F^{\mu\nu} \right) = 0 .
\label{sec2-40a}
\end{equation}
Comparing Eq. \eqref{sec2-40a} with the Maxwell + dilaton equation 
Eq. \eqref{sec2-20} we see that the dielectric permeability $\varepsilon$ 
can be connected with the dilaton field $\phi$ by the following way 
\begin{equation}
  \varepsilon = e^{-2\phi}
\label{sec2-41}
\end{equation}
and the electric displacement $\vec{D}$ 
\begin{equation}
  \vec{D} = \varepsilon \vec{E} = e^{-2\phi}\vec{E} = 
  \vec{E} + \vec{P} = 
  \left( 1 + 4\pi \chi \right)\vec{E}
\label{sec2-42}
\end{equation}
where $\vec{P}$ is the polarization vector. 
\par 
We would like to consider the static spherically symmetric case with the metric 
\begin{eqnarray}
  ds^2 & = & A^2(r) dt^2 - \frac{dr^2}{A^2(r)} - \rho^2(r) 
  \left( d\theta^2 + \sin^2\theta d\varphi^2\right),
\label{sec2-50}\\
  F_{tr} & = & E_r \neq 0.
\label{sec2-60}  
\end{eqnarray}
The solution is \cite{gregory}
\begin{eqnarray}
  A^2(r) & = & 1 - \frac{r_+}{r},
\label{sec2-65}\\
  \rho^2(r) & = & r(r - 2r_0),
\label{sec2-70}\\
  e^{2\phi(r)} & = & \frac{\rho^2}{r^2} = 
  1 - \frac{2r_0}{r},
\label{sec2-80}  \\
  E_r(r) & = & \frac{Q}{r^2},
\label{sec2-90}\\  
  D_r(r) & = & \frac{Q}{\rho^2},
\label{sec2-100}  \\
  r_0 & = & \frac{Q^2}{r_+}
\label{sec2-110}  
\end{eqnarray}
where $r_+$ and $r_0$ are some constants, $Q$ is the electric charge. 
For us interesting is the following case 
\begin{equation}
  r_0 \gg r_+ \qquad \text{or} \qquad r_+ \ll Q
\label{sec2-120}
\end{equation}
where the electric charge and field is in the natural units : 
$[Q] = cm$ and $[E] = cm^{-1}$. It is necessary 
to note that we have a gravitational singularity at the point 
$r = 2r_0$ ($\rho = 0$). 
\par
Our interpretation of this solution is the following. It is a bare electric 
charge $Q$ surrounded by a polarized spacetime foam (PSF). The PSF is described 
by the dilaton field $\phi$ which  is connected with the operator 
$W(x,y)$ describing the minimalist wormhole \eqref{sec1-30}. At the infinity 
$\phi(r) = 0$ that means $W(x,y) =0$ and $\varepsilon = e^{-2\phi} = 1$, 
\textit{i.e.} the PSF absents at the infinity.
Let us consider 
\begin{equation}
  W^{ab}(\vec x,\vec y) = \frac{1}{4} \ln 
  \left( 1 - \frac{2r_0}{\left| \vec{x} - \vec{x_0} \right|} \right)
  \left( 1 - \frac{2r_0}{\left| \vec{y} - \vec{y_0} \right|} \right)
  \theta^a \theta^b 
  \quad r \geq 2 r_0
\label{sec2-121}
\end{equation}
where $\phi (\vec x) = \phi (| \vec{x} - \vec{x_0}|)$; $\vec x_0$ 
and $\vec y_0$ 
are the locations of the mouthes of the minimalist wormhole (quantum handle). 
We see that $W^{ab}(\vec x,\vec y) \approx 0$ by 
$|\vec{x}_0 - \vec{y}_0| \gg 2 r_0$. It allows us to say that in the spacetime 
foam only the points on the distance $l \approx r_0$ can be pasted together 
making the quantum handle. In such interpretation $r_0 \approx l_{Pl}$. 
Nevertheless, in some external field the magnitude $r_0$ can have various 
values. As it was mentioned in Sec. \ref{model} the longitudinal size 
of the solution presented there can have arbitrary value which depends on the 
relation between electric $E$ and magnetic $H$ fields. It means that in the 
presence of a big external magnetic field the magnitude $r_0$ can be very 
large. 
\par 
Now we would like to present some physical consequences of our model of the 
bare charge imbedded in the PSF. 

\subsection{The screening of the bare charge}

In this subsection we would like to calculate a surface density of 
bounded electric charges connected with the polarization of the spacetime 
foam. At first we will calculate the normal component $P_n = P_r$ of the 
polarization vector 
\begin{equation}
  4\pi P_r = D_r - E_r = Q \left( \frac{1}{\rho^2} - \frac{1}{r^2} \right).
\label{sec2-130}
\end{equation}
The surface density $\sigma$ of the bound charges is 
\begin{equation}
  \sigma = P_n = P_r .
\label{sec2-140}
\end{equation}
Consequently a charge $Q_{anti}$ on the little sphere surrounding the bare 
charge $Q$ is 
\begin{equation}
  Q_{anti} = -\left| 4\pi \rho^2 \sigma \right| = 
  -Q \left( 1 - \frac{\rho^2}{r^2} \right) = -Q \frac{2r_0}{r} 
  \xrightarrow{r \rightarrow 2r_0} -Q .
\label{sec2-150}
\end{equation}
Thus the PSF completely shield the bare electric charge. Another consequence is that 
the PSF intercepts all the force lines of the bare charge and redistributes 
their in the surrounding space. 

\subsection{The energy of the shielded charge}

From the above-mentioned results we see that the spacetime foam surrounding 
the bare electric charge redistributes electric force lines by such a way that 
the charge is screened. This leads to the fact that the effective electric 
displacement can be introduced and as the consequence the energy density of the 
bare charge is changed. In this subsection we would like to calculate the 
energy of the electric charge surrounded by the PSF. 
Let us calculate the energy $W$ of the electric field 
\begin{equation}
  W = \frac{1}{8\pi} \int_{2r_0}^\infty D_r E_r \sqrt{-\gamma} dr 
  \approx \frac{Q^2}{4r_0}
\label{sec2-160}
\end{equation}
where we have used the condition $r_+ \ll r_0$ and $\gamma$ is the determinant 
of the 3D metric. Eq. \eqref{sec2-160} is very interesting. We see that 
the energy of the charge surrounded by the PSF depends on some characteristic length 
$r_0$  which is connected with the typical length of quantum handles in spacetime 
foam. Usually the typical length of quantum handles is $l_{Pl}$. In this case 
$W \approx Q^2/l^2_{Pl}$ and the advantage of this formula is evident : it is 
a finite number whereas the energy of the point charge in the classical 
electrodynamics is infinite number. 
\par 
As indicated above there is the possibility when the typical length of quantum handles 
can be larger (in the presence of a big external magnetic field). It is 
interesting to determine the value of $r_0$ so that this energy can be compared 
with the rest energy of electron 
\begin{equation}
  m_0 c^2 = \frac{e^2}{4r_0}, \qquad Q = e
\label{sec2-170}
\end{equation}
where $m_0$ is the rest mass of electron. In this case 
the typical length of quantum handles is $r_0 \approx 10^{-13}cm$ 
($r_+ \approx 10^{-45}$) and it can be compared with the classical radius 
of electron. Of coarse such value of quantum handles is too much. 

\subsection{The gravitational effects}

Now we would like to calculate some gravitational effects connected 
with the metric \eqref{sec2-50}. The metric approximately is
\begin{equation}
  ds^2 \approx dt^2 - dr^2 - r(r-2r_0)
  \left( d\theta^2 + \sin^2 \theta d\varphi^2 \right), 
  \quad r \geq 2r_0.
\label{sec2-180}
\end{equation}
We see that the metric in this model is not flat. The difference 
can be measured as a deficit of the sphere area 
\begin{equation}
  \delta = 1 - \frac{4\pi l^2_\rho}{4\pi \rho^2}
\label{sec2-190}
\end{equation}
where $l_\rho$ is the distance between the origin ($r=2r_0$) and the sphere with 
the radius $\rho = \sqrt{r(r - 2r_0)}$ 
\begin{equation}
  l_\rho = \int_{2r_0}^r \frac{dr}{\sqrt{1 - \frac{r_+}{r}}} 
  \approx \left( r-2r_0 \right).
\label{sec2-200}
\end{equation}
It means that 
\begin{equation}
  \delta \approx 1 - \frac{\left( r - 2r_0 \right)^2}{\rho^2} = 
  \frac{2}{3} \qquad \text{by} \qquad r = 3r_0.
\label{sec2-210}
\end{equation}
Let us remind that in the chosen coordinate system $r > 2r_0$ . 
\par 
Thus this model of electron with the PSF give us the finite rest energy of 
electron (connected with its renormalized electric field) but at the center 
of electron we have a gravitational singularity. Also we should note that at 
the center of electron the electric field $E_r = \frac{Q}{2r_0}$  is finite 
but the direction of this vector is undefined. The picture of $\vec{E}$ at the 
center is like to a hedgehog. 
\par 
The experimental verified consequences of this model is the nonflat metric 
\eqref{sec2-180} and the difference between $D_r$ and $E_r$. For example, at 
the distance $r - 2r_0 \approx r_0$ we have the deficit of the sphere area
\begin{equation}
  \delta \approx \frac{2}{3}
\label{sec2-220}
\end{equation}
and the dielectric permeability 
\begin{equation}
  \varepsilon = \frac{D_r}{E_r} = \frac{r^2}{\rho^2} = 
  \frac{1}{1 - \frac{2r_0}{r}} \approx 3.
\label{sec2-230}
\end{equation}
Evidently it is too much but it can be connected with that our dilaton model 
of the PSF is approximate one. Nevertheless on the basis of such model we have 
the interesting model of screening of the bare electric charge with 
the finite energy of the electric field. 

\section{The spinor field}

In this section we would like to consider the spinor model of the PSF with 
the following Lagrangian
\begin{eqnarray}
&& {\mathcal L} = \sqrt{-g}
\left \{
-\frac{1}{2k}
        \left (
        R + \frac{1}{4}\mathcal F_{\alpha\beta}\mathcal F^{\alpha\beta}
        \right ) +
\right.
\nonumber \\
&&
\left.
\frac{\hbar c}{2}
        \left [
        i\bar \psi
                \left (
                \gamma^\mu \nabla_\mu - 
                \frac{1}{8}\mathcal F_{\alpha\beta}
                \gamma^{[\alpha}\gamma^{\beta ]} -
                \frac{1}{4}l^2_0
                \left (\gamma^5\gamma^\mu \right )
                \left (i\bar\psi \gamma^5\gamma_\mu
                \psi \right ) - \frac{m}{i}
                \right )\psi  + h.c.
        \right ]
\right \}
\label{md2}
\end{eqnarray}
where $\mathcal F_{\alpha\beta} = \frac{c^2}{k} F_{\alpha\beta}$; 
$\mathcal F_{\alpha\beta}$ and $F_{\alpha\beta}$ are the Maxwell tensor 
in the natural and CGS units respectively;  
$\alpha ,\beta ,\mu$ are the 4D world indexes,
$\gamma^\mu$ are the 4D $\gamma$ matrixes with usual definitions
$\gamma^\mu\gamma^\nu + \gamma^\nu\gamma^\mu =
2\eta^{\mu\nu}$, $\eta^{\mu\nu} = (+,-,-,-)$ is the signature of the
4D metric; $l_0$ is some length; 
$k = 6.67 \times 10^{-8} cm^3g^{-1}s^{-2}$ is the gravitational constant. 
Of coarse Lagrangian \eqref{md2} is an arbitrary choice of a 
``fermion-generated'' spacetime foam model. Nevertheless there are some 
arguments for the nonminimal 
$(\bar{\psi} \mathcal F_{\alpha\beta} \gamma^\alpha \gamma^\beta \psi)$ 
and nonlinear 
$(\bar{\psi} \gamma^5 \gamma^\mu \psi) (\bar{\psi} \gamma^5 \gamma_\mu \psi)$ 
terms. These terms have the natural origin in the 5D gravitational Lagrangian with 
a torsion after the 4D dimensional reduction. Our hope is that the future 
investigations will give us the more natural explanation for this choice. 
Varying with respect to $g_{\mu\nu}$, $\bar\psi$
and $A_\mu$ leads to the following equations
\begin{eqnarray}
R_{\mu\nu} - \frac{1}{2}g_{\mu\nu}R & = & \varkappa T_{\mu\nu},
\label{md4}\\
D_\nu \mathcal H^{\mu\nu} & = & 0 ,
\,
\label{md5}\\
  \left[
  i\gamma^\mu \nabla _\mu - 
  \frac{1}{8} \mathcal F_{\alpha\beta}
  \left (
     i\gamma^{[\alpha} \gamma^{\beta ]}
  \right ) - 
  \frac{1}{2} l^2_0
  \left (
     i\gamma^5\gamma^\mu 
  \right )
  \left (
     i\bar\psi \gamma^5\gamma_\mu
  \right ) - m 
  \right]\psi &= &0 ,
\label{md6}\\
\mathcal H^{\mu\nu} & = & \mathcal F^{\mu\nu} + f^{\mu\nu} ,
\label{md6a}\\
f^{\mu\nu} & = & 4l^2_0
\left (
i\bar\psi \gamma^{[\mu} \gamma^{\nu ]}\psi
 \right ) 
\label{md7}
\end{eqnarray}
where $\nabla _\mu = \partial _\mu - \frac{1}{4} \omega _{ab\mu}
\gamma^{[a} \gamma^{b]}$ is the 4D covariant derivative of the spinor field,
$\omega _{ab\mu}$ is the 4D Ricci coefficients, $a,b$ are the fier-bein indices, 
$f_{\mu\nu}$ is the polarization tensor of the spacetime foam, 
$[]$ means the antisymmetrization. 
This equation set is very complicated therefore we 
will consider here only the PSF + electrodynamics, \textit{i.e.} gravity will 
be excluded.
\par 
Let us consider the case when we do not have the external electrical charges. 
It allows us to resolve the Maxwell equation \eqref{md5} 
\begin{equation}
  \mathcal F_{\mu\nu} = -f_{\mu\nu} = -4l_0^2 
  \left (
i\bar\psi \gamma_{[\mu} \gamma_{\nu ]}\psi
 \right ).
\label{sec2-240}
\end{equation}
Then the Dirac equation has the following form 
\begin{equation}
  \left\{
  i\gamma^\mu \nabla _\mu + \frac{l_0^2}{2}
   \left[
    \left(
    i\gamma^{[\mu}\gamma^{\nu ]}
    \right)
    \left(
    i\bar{\psi}\gamma_{[\mu}\gamma_{\nu ]}\psi
    \right) - 
   \left(
    i\gamma^5\gamma^\mu
    \right)
    \left(
    i\bar{\psi}\gamma^5\gamma_\mu\psi
    \right)
   \right] - m 
  \right\} \psi = 0.
\label{sec2-250}
\end{equation}
Immediately we see that it is one of variants of the non-linear Heisenberg 
equations (the Dirac equation + a non-linear term). The solution we will search 
in the following form 
\begin{equation}
  \psi (r, \theta , \varphi ) = e^{i\omega t }
  \begin{pmatrix}
    f(r) \\
    0    \\
    ig(r) \cos\theta\\
    ig(r) \sin\theta e^{i\varphi}
  \end{pmatrix}.
\label{sec2-260}
\end{equation}
After substitution in Eq. \eqref{sec2-250} we have 
\begin{eqnarray}
  g' + \frac{2}{r}g + f(m + \omega ) - l_0^2 
  f\left( f^2 - g^2 \right) & = & 0,
\label{sec2-270}\\
  f' + g \left( m - \omega \right) - l_0^2
  g\left( f^2 - g^2 \right) & = & 0.
\label{sec2-280}
\end{eqnarray}
In the Ref. \cite{finkel} it is shown that such equations set has regular 
solutions. At the origin $r = 0$ this solution can be expanded into a series
\begin{eqnarray}
  f(r) & = & f_0 + \frac{f_2}{2} r^2 + \ldots ,
\label{sec2-290}\\
  g(r) & = & g_1 r + \frac{g_3}{3!} r^3 + \ldots
\label{sec2300}
\end{eqnarray}
with 
\begin{eqnarray}
  g_1 & = & \frac{1}{3} f_* 
  \left[ l_0^2 f_*^2 - \left( m + \omega \right)
  \right], 
\label{sec2-310}\\
  f_2 & = & g_1 
  \left[ 
  l_0^2 f_*^2 - \left( m - \omega \right)
  \right]
\label{sec2-320}
\end{eqnarray}
where $f_*$ is a value of $f_0$ for which there is the above-mentioned regular 
solution for $f(r)$ and $g(r)$, \textit{i.e.} for the another 
$f_0 \neq f_*$ the solution has an infinite energy.
\par 
The asymptotical behavior of the regular solution is 
\begin{eqnarray}
  f(r) & = & \frac{\tilde{f}}{l_0 \sqrt{m}}
  \frac{e^{-r\sqrt{m^2 - \omega^2}}}{r} + 
  \ldots ,
\label{sec2-330}\\
  g(r) & = & \frac{\tilde{f}}{l_0 \sqrt{m}}\sqrt{\frac{m + \omega}{m - \omega}} 
  \frac{e^{-r\sqrt{m^2 - \omega^2}}}{r} + 
  \ldots \qquad .
\label{sec2-340}
\end{eqnarray}
where $\tilde{f}$ is some constant. 
Let us come back to the Maxwell fields. For our spinor ansatz we have the following 
components of the electric and magnetic fields 
\begin{eqnarray}
  \mathcal E^r(r) & = & -f^{tr}(r) = 8l_0^2 f(r) g(r),
\label{sec2-390}\\
  \mathcal H_\theta (r)& = & \frac{1}{2} \epsilon_{\theta r \varphi} f^{r \varphi}(r)
  = 4l_0^2 \sin\theta \Bigl( f^2(r) + g^2(r) \Bigr),  
\label{sec2-400}\\
  \mathcal H_r (r)& = & \frac{1}{2} \epsilon_{r\theta\varphi} \sqrt{-\gamma} 
  f^{\theta\varphi}(r) = 4 l_0^2 \cos\theta \Bigl( g^2(r) - f^2(r) \Bigr)
\label{sec2-410}  
\end{eqnarray}
here $\mathcal E_i$ and $\mathcal H_i$ are the electric and magnetic 
fields in the natural units. The corresponding magnitudes in the CGS units are 
$E_i = \frac{\sqrt{k}}{c^2} \mathcal E_i$ and 
$H_i = \frac{\sqrt{k}}{c^2} \mathcal H_i$. 
Now we have to do the following important simplifying remark. 
The ansatz for the spinor field \eqref{sec2-260} has a preferred direction (the 
spin direction) but the PSF is a quantum fluctuating object therefore we should 
average our results over the spin direction. It means that we should average 
electric and magnetic fields over $\theta$ ($0 \leq \theta \leq 2\pi$) 
that give us 
\begin{eqnarray}
  \mathcal H_\theta & = & \mathcal H_r = 0,
\label{sec2-420}\\
  \mathcal E_r & = & 8 l_0^2 f(r) g(r).
\label{sec2-430}
\end{eqnarray}
Now we are ready to formulate the result. Our solution with $\psi$  and 
$\mathcal E_r$ describes a ball of the polarized spacetime foam filled with the 
electric field. Another words, the PSF can confine electric field 
in some volume and it is \textbf{\textit{a pure quantum gravity phenomenon}}. 
We can expect that 
the energy density of the electric field will be very big and it will 
depend on the properties of the PSF, \textit{i.e.} on the typical length 
of quantum handles. 
\par 
Eq's \eqref{sec2-330} - \eqref{sec2-340} show us that the spinor field 
is nonzero inside of a region with the linear sizes 
$\approx r_0 = (m^2 - \omega^2)^{-1/2}$. It means that the operator 
$W(x,y)$ is nonzero in this region and consequently the typical length 
of quantum handles are $r_0$. 

\subsection{The energy of the PSF ball}

\begin{figure}
\begin{center}
\framebox{
\includegraphics[height=10cm,width=8cm]{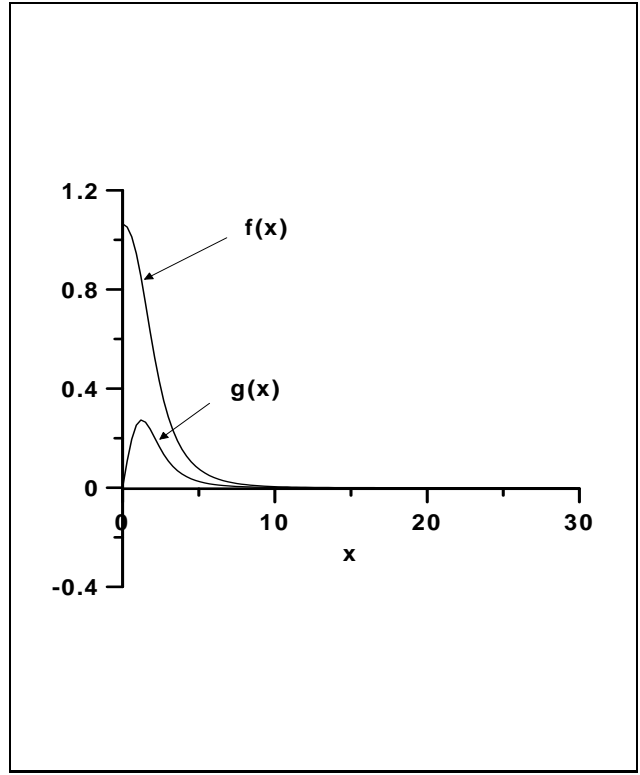}}
\vspace{5mm}
\caption{The functions $\bar{f}(x)$ and $\bar{g}(x)$. 
$\beta = 0.9$, $f_* \approx 1.06477126$.}
\label{fig2}
\end{center}
\end{figure}

The basic purpose of this section is to calculate the PSF ball with the 
confined electric field and compare this energy with some big interesting 
energies (nuclear bomb, Gamma Ray Burst (GRB)). 
At first we would like to write Eq's \eqref{sec2-270} \eqref{sec2-280} in 
the dimensionless form 
\begin{eqnarray}
  \bar g' + \frac{2}{x}\bar g + \bar f(1 - \beta) - 
  \bar{f}\left( {\bar f}^2 - {\bar g}^2 \right) & = & 0,
\label{sec3-10}\\
  \bar f' + \bar g(1 + \beta) - \bar{g}\left( {\bar f}^2 - {\bar g^2} \right) 
  & = & 0,
\label{sec3-20}\\
  \bar f(x) = \frac{l_0}{\sqrt{m}}f(r),
  \qquad
  \bar g(x) = \frac{l_0}{\sqrt{m}}g(r),
  \qquad
  x = r m ,
  \qquad 
  \beta & = & -\frac{\omega}{m}.
\label{3-40}  
\end{eqnarray}
The numerical solution of these equations presented on the Fig.\ref{fig2}.
\par
And now we can calculate the energy of the electric field 
\begin{equation}
  W = \frac{1}{8\pi}\frac{c^4}{k}
  \int_0^\infty \mathcal E_r^2 4\pi r^2 dr = 
  32 \frac{c^4}{k} \frac{1}{m} \int_0^\infty
  \bar{f}^2(x) \bar{g}^2(x) x^2 dx
\label{sec3-50}
\end{equation}
where the numerical coefficient $\frac{c^4}{k}$ is necessary for the conversion 
the natural units to the CGS units. The numerical value of the integral is of 
the order of 1. Therefore we can estimate the value of $m$ 
\begin{equation}
  32 \frac{c^4}{k}\frac{1}{m} \approx W 
  \quad \text{and} \quad 
  m \approx 32 \frac{c^4}{k} \frac{1}{W}
\label{sec3-60}
\end{equation}
Now we would like to calculate the radius $r_0$ and electric field $E$ 
for two characteristic energies. 
\par 
\textbf{1. Nuclear bomb}. In this case $W \approx 10^{20}erg$ and 
\begin{equation}
  m \approx 4 \times 10^{30} cm^{-1}
  \qquad \text{and} \qquad
  r_0 \approx \frac{1}{\sqrt{m^2 - \omega^2}} = 
  \frac{1}{m \sqrt{1 - \beta^2}}
  \approx 0.6 \times 10^{-30} cm
\label{sec3-61}
\end{equation}
The value of the electric field can be estimated as 
\begin{equation}
  E_r = \frac{c^2}{\sqrt{k}} 8l_0^2 fg = \frac{c^2}{\sqrt{k}} m 
  \bar{f} \bar{g} \approx 
  256 \frac{c^6}{k^{3/2}} \frac{1}{W}
\label{sec3-62}
\end{equation}
The values of the $\bar{f}$ and $\bar{g}$ inside of the ball are of the order 
of unity. It gives us 
\begin{equation}
  E_r \approx 10^{56} CGS
\label{sec3-63}
\end{equation}
Consequently the PSF ball filled with the electric field can contain the energy 
of the order of the energy of the nuclear bomb. It can happen if the typical 
length of the quantum handles is $\approx 10^{-30}cm$. 
\par 
\textbf{1. Gamma Ray Burst}. In this case $W \approx 10^{53}erg$ and 
\begin{equation}
  m \approx 4 \times 10^{-3} cm^{-1}
  \qquad \text{and} \qquad
  r_0 \approx \frac{1}{\sqrt{m^2 - \omega^2}} 
  \approx 0.5\times 10^{3} cm = 5.00 m
\label{sec3-70}
\end{equation}
The electric field is estimated as 
\begin{equation}
  E_r \approx 10^{23} CGS .
\label{sec3-90}
\end{equation}
Consequently the PSF ball filled with the electric field can contain the energy 
of the order of the energy of the GRB and the linear sizes of this ball are of the 
order of $\approx 10 m$. 
It means that : in order to obtain such energy $W \approx 10^{53} erg$ the 
typical length of quantum handles must be $r_0 \approx 10 m$. 
Again we would like remind that there is a possibility to make handles with 
the big longitudinal size : it can be in the presence of the big external 
magnetic field. In conclusion we would like to say that the big E\&M fields 
can produce quantum gravitational effects with the release of enormous 
amount of energy. 
\par
The value of such energies latent in the PSF ball 
is not surprising because it is quantum gravity effects. 
The big question here is : can be such energy of the electric field 
extracted from the frozen state in the polarized spacetime foam ? Another questions 
are : how can be such objects created and how fast it will disintegrate~?
Another interesting peculiarity of the spinor equation is that $l_0$ can be 
the Planck length $l_0 = l_{Pl}$ and nevertheless this microscopical 
equation can give the macroscopical energy and linear sizes ! 

\subsection{Non-linear Heisenberg equation}

It is interesting to note that Eq. \eqref{sec2-250} is one of the variants of 
the Heisenberg equation for a non-linear spinor field \cite{heis1}. This remark 
allows us to take a new interpretation of the solutions obtained in these 
Ref's : it is the PSF ball on a microscopical level. 

\section{Conclusions}

In this paper we offer an approximate model of the polarized spacetime foam. 
The microscopically description is based on the operator $W(x,y)$ of 
creation/destruction the minimalist wormholes. The minimalist wormhole is some 
approximation for the quantum handle when it is contracted 
into a point. The model for the quantum handle is given by the wormhole-like 
solution of the 5D Kaluza-Klein gravity. This metric contains off-diagonal 
components of the 5D metric $G_{5t}$ and $G_{5\varphi}$ which can be considered 
as an electric and magnetic fields. 
It allows us to consider each quantum handle (minimalist wormhole) as an electric 
dipole. Obviously that in such point of view the spacetime foam can be polarized 
in the presence of an external electric field. 
\par 
The operator $W(x,y)$ can be connected with either a scalar or a spinor field. 
Here the question arises : is this operator a consequence of quantum gravity 
or it is an additional stuff in quantum gravity describing the topology 
changes ? 
We assume that these fields can be dynamical. As a model of the scalar field 
we propose the dilaton field. In this case the dilaton field describes 
a polarization 
of the spacetime foam. In the result we have a shielding of the bare electric 
charge and the finiteness of the energy of the electric field. The physical 
sense of the scalar/spinor fields in the operator $W(x,y)$ is the following : 
\textbf{\textit{the product $\varphi(x) \varphi(y)$ (or $\psi_\alpha(x) 
\psi_\beta(y)$) 
describes the typical length of quantum handles in the spacetime foam}}, 
\textit{i.e.} the linear sizes of the region where this product is nonzero 
characterizes the typical length of quantum handles. 
\par 
In the another variant a spinor field describes the polarization of the 
spacetime foam. We have considered the simplest case without external electric 
charges and gravity. We have shown that the polarized spacetime foam can confine 
an electric 
field in some finite region of the space. The magnitude of the confined electric 
field can be very big. We have shown that this energy depends on the 
typical length of quantum handles and for the energy of nuclear bomb this 
length is $r_0 \approx 10^{-30} cm$ and for the GRB it is 
$r_0 \approx 10 m$. 
It is necessary to note that in the consequences 
of the very high energy density of the electric field confined in the PSF 
the experimental verification of the effects connected with quantum gravity is 
very dangerous, \textbf{\textit{much more dangerous then the experiments with the 
nuclear energies !}} 
\par 
In this paper we have considered two possibilities for the description 
of the spacetime foam : the scalar and spinor fields. Of coarse, the choice 
between these two possibilities can be made either on the basis of an exact 
theory of quantum gravity or from an experiment. Unfortunately, we do not 
have neither. Therefore we can only compare the consequences of two 
approaches and choose more natural. The comparison of solutions 
for the scalar field Eq's \eqref{sec2-65} - \eqref{sec2-100} and spinor field 
Fig. \ref{fig2} shows us that the first solution has a singularity at the 
center of the field whereas the spinor field does not have a singularity. 
It means that the spinor field is more natural choice for the description 
of the PSF. Another argument for this statement is the following. Let us 
compare the polarization vector for the ``scalar-generated'' model of the 
spacetime foam 
\begin{equation}
  \mathcal{P}_i = \left( e^{-2 \phi} - 1\right) E_i 
\label{concl-10}
\end{equation}
and for the ``fermion-generated'' model of the spacetime foam 
\begin{equation}
  \mathcal{P}^i = f^{ti} = 4l_0^2 
  \left( i \bar{\psi} \gamma^{[t} \gamma^{i]} \psi \right)
\label{concl-20}
\end{equation}
Now we see how is the difference between the ``scalar-generated'' and 
``fermion-generated'' models of the spacetime foam. For the first case the 
polarization vector \eqref{concl-10} is proportional to an external electric 
field $E_i$ whereas in the second case $\mathcal P_i$ does not direct depend 
on this field. The second case is more interesting since in this case we have 
very intriguing situation : an electric field can be confined with the 
spacetime foam in some finite region without any hairs at the infinity. 
In fact such situation was discussed in Sec.5. It is very unusual situation 
when a static electric field exists without any electric charges. The 
explanation for this fact is very simple : the sum of this electric field 
and the electric field generated by spacetime foam is zero 
(see, Eq. \eqref{sec2-240}). 
\par 
Finally we would like to summarize
\begin{itemize}
  \item 
  Each quantum handle in the spacetime foam is like to an electric dipole. 
  \item 
  It is possible to introduce an operator describing a quantum handle. 
  \item 
  In the presence of an external electric field the spacetime foam can be polarized. 
  \item 
  The polarized spacetime foam can shield a bare electric charge.
  \item 
  The energy of the shielded charge becomes finite.  
  \item 
  The polarized spacetime foam can confine electric field into finite region. 
  \item 
  The energy of confined electric field can be very big.
\end{itemize}
The problems originating here are the following
\begin{itemize}
  \item 
  At the center of the shielded charge there is a gravitational singularity. 
  It is possible to avoid it in some more realistic scenario ?
  \item 
  Must be quantized the scalar and spinor fields ? Is it not trivial question because 
  these fields are some 
  approximate description of the spacetime foam and consequently they already 
  connected with quantum gravity. 
  \item
  How can be created the ball with the confined electric fields : is it a  
  quantum fluctuation of the metric or it was created in the Early Universe 
  and now only disintegrate ?
  \item 
   What is the duration of the life of these objects : exist they infinitely or 
   disintegrate after some finite time ?
  \item
  What give us switching on the gravity : the ball will be the same with some 
  modifications or something different~?
\end{itemize}

\section{Acknowledgment}
I am very grateful for Viktor Gurovich for the fruitful discussion, 
ISTC grant KR-677 for the financial support and the Alexander von Humboldt 
Foundation for the support of this work.

\end{document}